# Near-field electrospinning of conjugated polymer light-emitting nanofibers

Daniela Di Camillo,[a,‡] Vito Fasano,[b,c,‡] Fabrizio Ruggieri,[a] Sandro Santucci,[a] Luca Lozzi,[a*] Andrea Camposeo[c,d*] and Dario Pisignano[b,c,d*]

[a] Dipartimento di Scienze Fisiche e Chimiche, Università degli Studi dell'Aquila, via Vetoio, I-67100 L'Aquila (Italy), e-mail: luca.lozzi@aquila.infn.it

[b] Dipartimento di Matematica e Fisica "Ennio De Giorgi", Università del Salento, via Arnesano I-73100 Lecce (Italy),

e-mail: dario.pisignano@unisalento.it

[c] Center for Biomolecular Nanotechnologies @UNILE, Istituto Italiano di Tecnologia (IIT), via Barsanti 1, I-73010 Arnesano, LE,(Italy),

[d] National NanotechnologyLaboratory of Istituto Nanoscienze-CNR, via Arnesano, I-73100 Lecce (Italy),e-mail: andrea.camposeo@nano.cnr.it

The authors report on the realization of ordered arrays of light-emitting conjugated polymer nanofibers by near-field electrospinning. The fibers, made by poly[2-methoxy-5-(2-ethylhexyloxy)-1,4-phenylenevinylene], have diameters of few hundreds of nanometers and emission peaked at 560 nm. The observed blue-shift compared to the emission from reference films is attributed to different polymer packing in the nanostructures. Optical confinement in the fibers is also analyzed through self-waveguided emission. These results open interesting perspectives for realizing complex and ordered architectures by light-emitting nanofibers, such as photonic circuits, and for the precise positioning and integration of conjugated polymer fibers into light-emitting devices.

**Introduction**

Polymer nanofibers are flexible 1-dimensional (1D) nanostructured materials, that are increasingly utilized in many fields such as tissue engineering,[1] filtration,[2] sensing,[3] and optoelectronics.[4] The decrease of the fiber characteristic dimensions below the range of few hundreds of nanometers may reveal new intriguing mechanical,[5] thermal[6] and optoelectronic properties,[4] mainly attributed to the particular packing and assembly of the polymer backbones at the nanoscale. The scientific and technological relevance of polymer nanofibers is stimulating a lot of research efforts to develop novel methods for the fabrication of 1D organic nanostructures with controlled shape, composition and assembly. Electrospinning (ES) is the most versatile and widely used technique for the fabrication of polymer nanofibers.[7-10] The elongation of a polymer solution by electrostatic field is the key feature of ES, easily allowing the fabrication of continuous polymeric and hybrid fibers with diameters down to few tens of nanometers. Typically, these nanostructures are collected in non-woven mats or uniaxially arrays,[11] and several regimes of jet instability can limit the precise positioning of individual fibers. This is particularly relevant when single active nanofibers have to be integrated into devices such as organic light-emitting diodes,[12] field-effect transistors[13] and optical sensors.[3] Recently, scanning tip ES[14] and near-field electrospinning (NF-ES)[15,16] have been demonstrated to be suitable for the synthesis and positioning of single nanofibers with high precision. In particular NF-ES, utilizing the stable region of the electrospun jet close to the metallic needle, has been exploited to fabricate polyethylene oxide (PEO),[15,16] poly(vinylidene fluoride),[17] polycaprolactone[18] and composite (PEO/carbon nanotubes)[19] fibers, with diameters down to 16 nm and assembled even in complex three-dimensional patterns.[20,21] $SnO_2$ and $TiO_2$ nanofibers have been also fabricated by depositing and annealing well-ordered polymeric fibers embedding suitable precursors.[22,23] However, the application of NF-ES to light-emitting conjugated polymers is still basically unexplored, notwithstanding the clear added value of these materials for a large variety of optoelectronic applications, many of which have already reached commercialization.

‡ These authors equally contributed to this work.

In this paper we report on light-emitting conjugated polymer nanofibers realized by NF-ES. Conjugated polymers cannot be easily electrospun due to their limited solubility, chain rigidity and low molar mass. Despite such difficulties, nanofibers made of these polymers have been produced by standard ES systems.[24-27] However the integration of individual light-emitting nanofibers in photonic and electronic devices is still difficult, due to the limited ability of standard ES systems to control the spatial deposition of individual polymer nanofibers between electrodes or other functional interfaces. NF-ES can largely outperform conventional ES for depositing conjugated polymer fibers with high spatial control. Moreover, NF-ES can enable also an improved control of fiber morphology,[15,20] that is essential for the exploitation of single active fibers for photonic or sensing applications. We realize well-aligned light-emitting fibers by the prototype conjugated polymer, poly[2-methoxy-5-(2-ethylhexyloxy)-1,4-phenylenevinylene] (MEH-PPV), producing ordered arrays. The fibers exhibit blue-shifted photoluminescence (PL) compared to reference films, and self-waveguiding of the emission. These results open the way for the fabrication of complex arrays and photonic circuits composed by nanofibers, exploiting the unique features of conjugated polymers for emitting, guiding and sensing UV-visible light.

**Experimental**

**Near-field electrospinning**

The light-emitting fibers are electrospun by mixing MEH-PPV with PEO, used as polymer carrier. We dissolve separately 250 mg of PEO (molecular weight, MW=300,000) in either 3.5 mL of a acetonitrile/toluene mixture (65/35 v/v) or 3 mL of acetic acid/toluene (17/83 v/v), and 10 mg of MEH-PPV (MW=380,000, Sigma Aldrich) in 2 mL of toluene. Among the different tested solvent combinations, those reported in the following allow the formation of a stable jet and uniform fibers to be obtained. Typically, fibers spun from acetic acid/toluene solutions feature a more uniform emission, whereas the presence of acetonitrile favors the formation of a higher number of fluorescent bead-like structures (see also below). The MEH-PPV solution is stirred for 4 hours until complete dissolution of the polymer. After stirring vigorously the PEO solution for 8 hours, 500 μL of MEH-PPV solution is added.

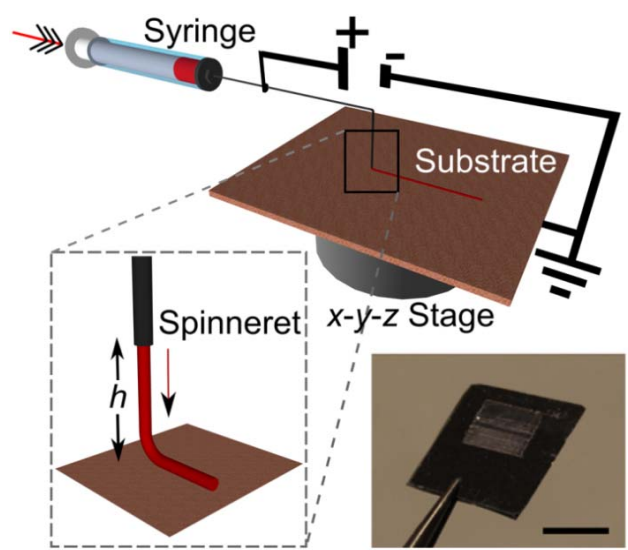

**Fig. 1** Schematics of the NF-ES system utilized for the realization of conjugated polymer nanofibers ($h$=500 μm). Inset: picture of a 5×5 mm$^2$ nanofibers array. Scalebar: 5 mm.

The resulting relative concentration of MEH-PPV/PEO is ~1:100 (w/w), and this solution is stirred and ultrasonically agitated, thus becoming homogeneous. SiO$_2$/Si (oxide thickness = 800 nm), cleaned with acetone in an ultrasonic bath and dried with pure air, is used as substrate. The substrate is placed on a grounded metallic plate (collector), 500 μm below the needle tip to utilize the stable liquid jet region for a controllable deposition of the fibers.

The aligned fibers are deposited by a home-built NF-ES apparatus, schematized in Figure 1. The NF-ES set-up is composed by a plastic syringe equipped with a μm-diameter tip Tungsten spinneret in a 26 gauge needle, a syringe pump (Harvard Apparatus 22), a high voltage power-supply (Innotec A2K5-20HR), a grounded collector and $x$-$y$ (C-865 PILine Ultrasonic Piezomotor Controller) and $z$ stages (C-862 Mercury DC-motor Controller), allowing to control the collector movement by dedicated software. The solution is loaded into the syringe, whose metallic needle is connected to the positive electrode of the power-supply and fed through the syringe needle at constant rate (50 μL/h). The applied electrostatic voltage is about 1.3 kV. During the deposition, the $x$-$y$ stage velocity is 50 cm/s. With these values of voltage and stage velocity, and at a tip-substrate distance of 500 μm, straight, continuous and uniform fibers are deposited. Lower stage velocities determine the formation of spiraling and buckling nanofibers, whereas non-continuous nanofibers would be deposited by using higher velocities. As a general rule, smooth and continuous fibers are formed when the stage translational velocity is as much as possible comparable to the deposition rate.[16] The experiments are carried out at room temperature and in air atmosphere. The morphology of the electrospun nanofibers is examined by Scanning Electron Microscopy (SEM), by using a FEG-SEM (LEO 1530) at 5-10 kV beam energy. Atomic force microscopy is performed by using a Multimode system equipped with a Nanoscope IIIa electronic controller (Veeco Instruments). Si cantilevers with a resonance frequency of 250 kHz are used to image the nanofiber surface topography in Tapping mode.

**Optical properties**

Fluorescence micrographs of the nanofibers are acquired by a laser confocal microscope, composed by a scan head (A1R MP, Nikon) and an inverted microscope (Eclipse Ti, Nikon). Samples are excited by a CW diode laser (Melles Griot 56 ICS series, $\lambda$ = 408 nm). The emission is collected by a 20× (Plan Fluor Numerical Aperture, $NA$=0.50, Nikon) or a 60× (oil immersion Plan Apo, $NA$=1.40, Nikon) objective and the intensity is measured by either a photomultiplier or a spectral detection unit equipped with a multi-anode photomultiplier (Nikon).

The single-nanofiber emission is analyzed by using a micro-photoluminescence ($\mu$-PL) set-up, based on an inverted microscope (IX71, Olympus) equipped with a 60× oil immersion objective (Plan ApoN, $NA$=1.42, Olympus). The PL is excited by a diode laser ($\lambda$ = 405 nm), coupled to the microscope by a dichroic mirror and focused on the sample through the microscope objective (spot diameter about 3 µm). The fiber emission is collected by the same microscope objective and analyzed by a monochromator (Jobin Yvon, iHR320), equipped with a Charged Coupled Device (CCD) detector (Jobin Yvon, Symphony). This $\mu$-PL system is also used for characterizing fiber waveguiding. Part of the light emitted by the fluorescent conjugated polymer, excited by the tightly focused laser beam, is coupled into the fiber and waveguided. The fiber optical losses can be evaluated by measuring the intensity of PL signal diffused by the fiber surface, as a function of the distance, $d$, from the exciting laser spot.[26] To this aim, the spatially-resolved emission intensity map of a single light-emitting fiber is measured by a CCD camera (Leica DFC 490).

For polarized excitation spectroscopy, the fibers are excited by a collimated, linearly polarized laser beam (diameter about 2 mm), normally incident on samples. The beam polarization can be adjusted with respect to the nanofiber longitudinal axis by a $\lambda/2$ waveplate. The nanofiber emission is collected by an optical fiber and analyzed by a spectrometer.

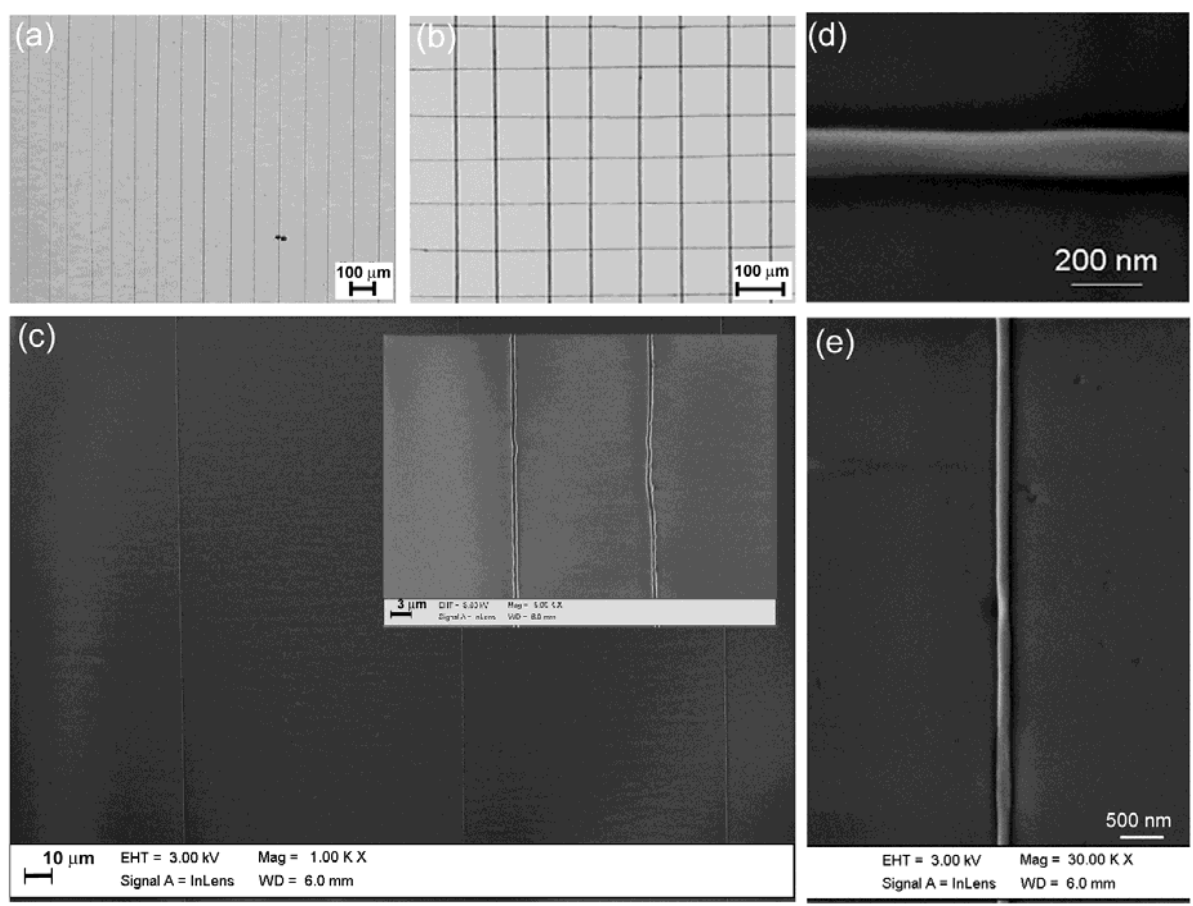

**Fig. 2** (a)-(b). Optical micrographs of ordered arrayed of MEH-PPV nanofibers. Scale bar: 100 μm. (c) SEM image of well-aligned fibers deposited by NF-ES. The distance between adjacent fibers is about 100 μm. Inset: high resolution SEM image. Scale bar: 3 μm. (d)-(e) SEM images of individual light-emitting nanofibers realized by NF-ES with diameter of the order of 100 nm.

## Results and Discussion

The key factor of NF-ES is the exploitation of the stable region of the extruded jet close to the metallic needle ($h$=500 μm in our experiments, Fig. 1). Such approach allows the onset of the jet instabilities to be avoided, and arrays of polymer nanofiber with high spatial precision to be consequently deposited. Figure 2(a) shows optical bright-field images of ordered arrays of MEH-PPV/PEO nanofibers, composed by parallel, roughly equally-spaced fibers (deposited area about 25 mm$^2$). The realized parallel fibers feature a very high degree of mutual alignment. The distribution of the angles formed by the fiber axes and the stage translational axis is characterized by a standard deviation<1°, much lower than the typical values reported in uniaxially aligned fibers realized by standard electrospinning systems.[28]

In Figure 2(b), two of such arrays have been sequentially deposited along perpendicular directions, evidencing the potentiality of the NF-ES technique for fabricating ordered arrays of conjugated polymer fibers with complex architectures. In Figure 2(c), we display SEM images, evidencing the very good alignment of fibers, which have a diameter down to 100 nm [Fig. 2(d)-(e)] nm and are spaced by about 100 μm. The nanofiber surface topography and height profile measured by atomic force microscopy (AFM) are shown in Fig. 3. Most of fibers have a ribbon shape, with height below 100 nm.

In Figure 4 we show PL micrographs of the produced MEH-PPV/PEO nanofibers, imaged by confocal microscopy. The data show uniform and almost regular arrays of fluorescent fibers, confirming the spatial control achieved in depositing electrospun fibers into pre-

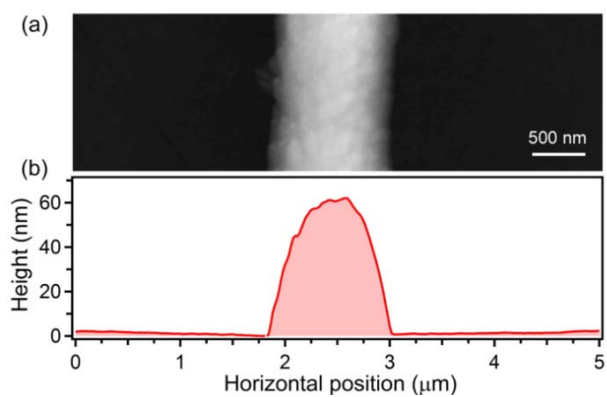

**Fig. 3** (a) Atomic force microscopy image and (b) height profile of a light-emitting fiber made by NF-ES.

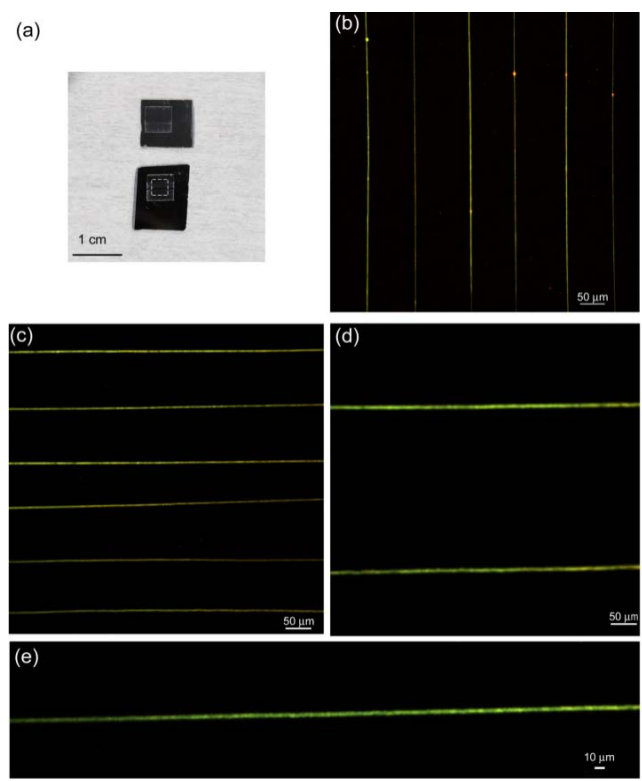

**Fig. 4** (a) Pictures of the fibers arrays realized by NF-ES. The dashed box highlights the region with nanofibers. (b)-(d) Confocal fluorescence micrograph of an array of emitting nanofibers made by NF-ES. (e) Single fiber confocal fluorescence image.

defined geometries, and the high effectiveness of the near-field technique for fabricating ordered arrays of emitting nanofibers. In some samples, high-resolution confocal maps of single fibers [Fig. 4(e)] evidence a complex structure internally to the electrospun nanostructures, with some bead-like features (linear density about 0.5 μm$^{-1}$) showing a more intense emission. We point out that these features can be observed only by confocal microscopy, that is sensitive to the active component of the fiber (MEH-PPV), whereas no internal structure is appreciable in fibers by SEM (Fig. 2). These features, observed also in other similar systems,[20] are likely due to aggregation and phase-separation effects occurring in the MEH-PPV/PEO blend.[29] The formation of electric field-induced internal microstructures, and the occurrence of phase separation phenomena as well as of superficial chemistry modifications have been reported also in polymer fibers produced by standard electrospinning systems, and exploited to tailor, for instance, the surface and mechanical properties of the fibers.[30-32]

The PL emission spectrum of the NF-ES fibers is peaked at about 560 nm with a full width at half maximum of about 90 nm, due to the overlap of different emission species and vibronic series [Fig. 5(a)]. For comparison, the broader and featureless emission spectrum of a reference film, drop-cast from the same polymer solution used for the NF-ES process [Fig. 5(a)], is peaked at about 586 nm and has a width of about 120 nm. We cannot rule out re-absorption effects occurring in the reference films, which would have as net effect the red-shift of the collected spectrum. Based on previous study of MEH-PPV films with various thickness,[33] we estimate that about 50% of the measured PL shift can be attributed to re-absorption effects.

However, the spectral shift measured here in this work between the emission from single fibers and from the reference film ($\cong$ 10 nm, taking into account the contribution of self-absorption) is larger than in previous studies.[33,34] The PL of MEH-PPV at the solid state has been widely investigated, and it can be attributed mainly to excitons, either localized on a single conjugated polymer chain (intrachain), or shared by neighboring polymer chains (interchain).[35-37] Compared to intra chain excitons, the emission from interchain species is typically red-shifted and characterized by a lower quantum yield. The formation of interchain excitons can be favored by aggregation effects, generally occurring in solid state samples or in highly concentrated solutions.[35,37] The observed blue-shift of the single fiber emission suggest a decrease of the emission from aggregates and interchain states in the nanofibers compared to the films, as observed also in MEH-PPV nanostructures produces by the standard electrospinning process.[33,34,38]

This effect can also be due to a reduced energy migration towards low energy sites. In fact, conjugated polymers are characterized by a broad absorption and emission spectrum, which is in part due to the inhomogeneous distribution of excited states energies, strongly correlated to the distribution of the conjugation lengths. The packing of the conjugated polymer backbones in different shapes breaks the conjugation length in sub-segments (conjugations lengths),[39] leading to a distribution of the excited states energies. Once the exciton is created by absorption of light, it can undergo electronic energy transfer towards the lower energy sites,[40,41] resulting in red-shifted emission.

This electronic energy migration depends also on sample morphology.[42,43] The observed large blue-shift could in part be due to a reduced energy migration in the nanofiber, that can be related to the molecular packing of the conjugated polymers into the fibers induced by near-field electrospinning. Several studies have evidenced the peculiar packing of the polymer macromolecules in electrospun nanofibers, induced by the applied electric field and by the stretching related to the high strain rate characteristic of the process.[31,32,44]

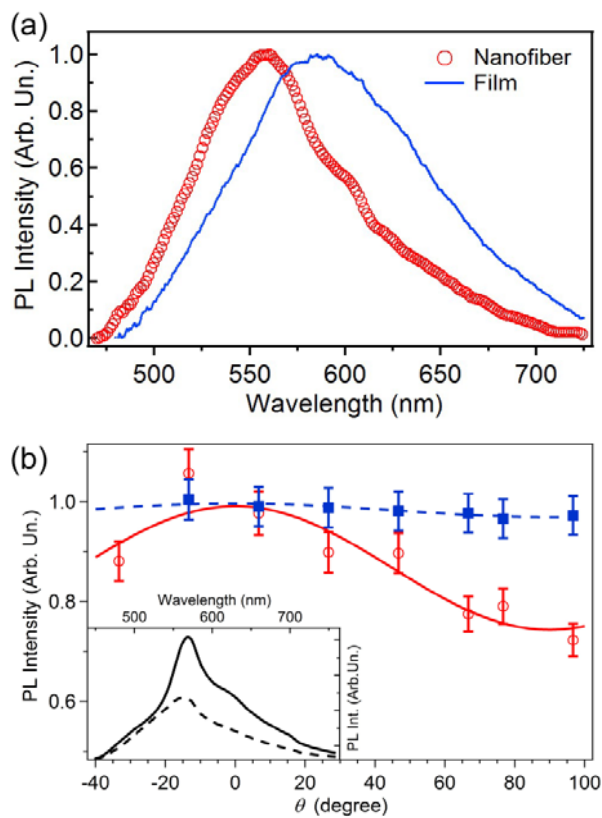

**Fig. 5** (a) μ-PL spectra of a single nanofiber (open circles) and of a reference film (continuous line). (b) PL intensity (empty circles) *vs* the incident laser polarization angle, *θ*, measured with respect to the fiber longitudinal axis (*θ*=0° for incident laser polarization parallel to the fiber axis, whereas *θ*=90° for incident laser polarization perpendicular to the fiber axis). Data obtained for a spincast reference film are also shown (full squares). Data are normalized to the maximum fluorescence intensity. Continuous and dashed lines are best fits to the data by a $\cos^2(\theta)$ expression. Inset: polarized emission spectra acquired with the analyzer axis parallel (continuous line) and perpendicular (dashed line) to the fiber axis. Fiber emission is excited with the excitation laser polarization parallel to the fiber axis. Similar measurements performed on spin-cast films show unpolarized emission.

The packing and possible alignment of MEH-PPV macromolecules within the electrospun nanofibers is investigated through polarized excitation and PL spectroscopies. The polarized excitation measurements are accomplished by collecting the nanofiber fluorescence intensity, $I(\theta)$, excited by a laser beam, whose linear polarization direction forms an angle, *θ*, with respect to the fiber axis. By this method the polarization dependence of the absorption process through the emission intensity can be probed.[45] Figure 5(b) shows the dependence of the fiber fluorescence intensity on *θ*, evidencing anisotropic polarized absorption and a predominant alignment of MEH-PPV optical transition dipoles along the nanofiber axis, since emission (absorption) is maximized for incident excitation polarization parallel to the fiber axis. For comparison, spincast films, where MEH-PPV molecules are expected to be randomly distributed, do not show any dependence of the emission intensity on incident polarization direction [Fig. 5(b)].

Interestingly, the measured fluorescence intensity modulation depths, $M = (I_{Max} - I_{Min}) / (I_{Max} + I_{Min})$ (where $I_{Max}$ and $I_{Min}$ are the maximum and minimum fluorescence intensity, respectively, obtained upon varying the excitation polarization direction), are in the range 0.1-0.3, close to values calculated for an ensemble of MEH-PPV molecules with uncollapsed, defect-coil conformation, whereas in ensembles of single molecules spincast from solution higher values are typically measured, that can be related to the presence of both collapsed and uncollapsed conformations.[39] These results are substantiated by the analysis of the fluorescence polarization [inset of Fig. 5(b)], evidencing that the emission is mainly polarized along the nanofiber axis (the peak emission with polarization parallel to the fiber axis is about twice the emission peak intensity with polarization perpendicular to the fiber axis), further supporting a prevalent orientation of the MEH-PPV optical transition dipoles along the fiber. The degree of ultimately achievable orientation of the active macromolecules in fibers produced by NF-ES is likely related to the solvent evaporation rate. Indeed, the alignment of MEH-PPV molecules could be adversely affected by the presence of residual solvent in the deposited fibers.[16] In fact, the short needle-collector distance can disfavor the complete evaporation of the solvents from the spun jet. As a consequence, the presence of solvent residues in the deposited fibers can allow the polymer molecules to still relax to reach a more isotropic configuration following deposition and during solidification,[32] thus decreasing the fluorescence degree of polarization. For light-emitting macromolecules, this eventual relaxational dynamics should also be related to the solvent quality, with good or poor solvents for the used conjugated polymer favoring typically more elongated or aggregated configurations, respectively.[36,37]

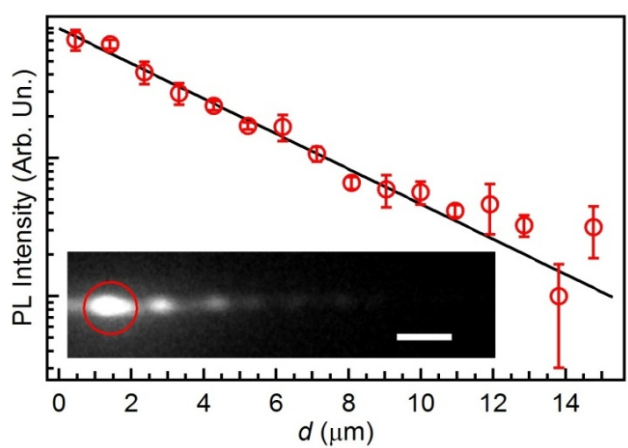

**Fig. 6** Normalized PL intensity guided by a single fiber as function of distance, $d$, from the excitation spot. The continuous line is a fit to the data by an exponential decay $I=I_0\exp(-\alpha d)$, where $\alpha$ is the optical loss coefficient. Inset: fluorescence micrograph of a MEH-PPV fiber. The PL is excited by a focused laser (red circle) and a part of emitted light is coupled into the fiber. Scale bar = 2 μm.

Finally, we investigate the waveguiding properties of the light-emitting nanofibers, in view of their possible exploitation as active waveguides. Figure 6 displays the spatial decay of the self-waveguided emission. The decay is almost exponential, with a loss coefficient of the order of $10^3$ cm$^{-1}$, comparable to similar active systems.[26,45] We attribute optical losses mainly to scattering from the surface fiber microstructure evidenced in Fig. 3. The measured surface roughness (root mean squared, RMS) is of about 10 nm (Fig. 3). Another contribution to the measured losses is attributable to self-absorption that typically affects waveguiding in conjugated polymer nanostructures and films. We anticipate that the optical losses can be reduced to cm$^{-1}$ level, suitable for on-chip waveguiding application, by using specifically-designed host-guest donor-acceptor systems.[46] These experiments are currently in progress in our laboratories.

## Conclusions

In summary, we demonstrate the possibility of realizing ordered arrays of light-emitting conjugated polymer fibers by NF-ES. We realize arrays of parallel and crossed fibers, composed by nanostructures with diameter of about 500 nm, emitting at 560 nm. The fibers show self-waveguiding of the emission with losses mainly limited by self-absorption and scattering from the fiber surface and internal nanostructure. The achieved high degree of control of fiber positioning would allow arrays of uniaxially aligned fibers to be realized, inheriting the anisotropic optical properties of single fibers. In perspective, the technique can be used to fabricate arrays of optically active nanofibers for sensing and photonic circuits applications and to deposit single active nanofiber in light-emitting devices.

## Acknowledgement

The research leading to these results has received funding from the European Research Council under the European Union's Seventh Framework Programme (FP/2007-2013)/ERC Grant Agreement n. 306357 (ERC Starting Grant "NANO-JETS").